\documentclass[journal]{IEEEtran} 
\usepackage{amssymb}
\usepackage{mathtools}
\usepackage{amsmath}
\usepackage{graphicx}
\usepackage{float}
\usepackage{epsfig}
\usepackage{csquotes}
\usepackage{caption}
\usepackage{subcaption}
\usepackage[T1]{fontenc}
\usepackage{color}
\usepackage{subcaption}
\usepackage{cite}
\usepackage{lettrine}
\usepackage{url}

\DeclareMathOperator*{\argmin}{arg\,min}

\begin{document}
\title{Determining the Dimension of the Improper Signal Subspace in Complex-Valued Data} 
\author{\IEEEauthorblockN{Tanuj Hasija,~\IEEEmembership{Student Member,~IEEE,} Christian Lameiro,~\IEEEmembership{Member,~IEEE,} and \newline Peter J. Schreier,~\IEEEmembership{Senior~Member,~IEEE}\\}
\thanks{ The authors are with the Signal and System Theory Group, Universit\"{a}t Paderborn 33098, Germany (e-mail: tanuj.hasija@sst.upb.de, christian.lameiro@sst.upb.de, peter.schreier@sst.upb.de).}
\thanks{This research was supported by the German Research Foundation (DFG) under grant SCHR 1384/3-1.}} 

\maketitle

\begin{abstract}
\textbf{
A complex-valued signal is improper if it is correlated with its complex conjugate. The dimension of the improper signal subspace, i.e., the number of improper components in a complex-valued measurement, is an important parameter and is unknown in most applications. In this letter, we introduce two approaches to estimate this dimension, one based on an information-theoretic criterion and one based on hypothesis testing. We also present reduced-rank versions of these approaches that work for scenarios where the number of observations is comparable to or even smaller than the dimension of the data.  Unlike other techniques for determining model orders, our techniques also work in the presence of additive colored noise. 
}
\end{abstract}


\section{Introduction}
\lettrine[findent=2pt]{{\textbf{A}}}{ }\textsc{Complex-valued} random vector $\mathbf{x}$ is proper if it is uncorrelated with its complex conjugate ${\mathbf{x}^\ast}$, and otherwise improper.  While propriety is a common assumption, improper signals arise in numerous areas in engineering such as communications and also in applied sciences such as oceanography and biomedicine \cite{schreier2010statistical,mooers1973technique, adali2007complex}. Detecting the number of improper signal components in a measurement is often a prerequisite before performing further steps like estimating the direction of arrival (in array processing) or blind source separation \cite{chevalier2007widely,roemer2009multidimensional,charge2001non, li2011application}. 

This detection problem can be solved as part of the more general problem of partitioning the observation space into signal and noise subspaces. The standard approach to achieve this partition is based on principal component analysis (PCA) and information theoretic criteria (ITC) \cite{wax1985detection}. However, 
this approach is suboptimal when some or all the signals in the observed data are improper. This is because this technique only takes into account the statistics of the covariance matrix $\mathbf{R}_{xx} = E[\mathbf{xx}^H]$ and ignores the complementary covariance matrix $\widetilde{\mathbf{R}}_{xx} = E[\mathbf{xx}^T]$. The covariance and complementary covariance matrices are both required to characterize the second-order characteristics of $\mathbf{x}$ \cite{schreier2010statistical}. 

Noncircular PCA (ncPCA) introduced in \cite{li2011noncircular} improves on PCA by also taking into account the information about impropriety contained in the complementary covariance matrix. Based on ITC, \cite{li2011noncircular} determines the dimensions of both the proper and improper signal subspaces from noisy observations. However, in some applications, we might only be  interested in the dimension of the improper subspace, for instance, when we know that all signal components are improper \cite{charge2001non}. This is the problem we solve in this letter. Even though the technique in \cite{li2011noncircular} can be used for this scenario as well, it is to be expected that a specialized technique works better than a more general one. Indeed, by determining the number of improper signal components only, we are able to reduce the number of required samples and relax the assumption on the noise structure. We only need to assume that the noise is proper, but unlike typical PCA-based methods, it does not have to be white.

We introduce two alternative approaches: one that is based on the minimum description length (MDL) ITC (see Section \ref{sec:MDL-ITC}), and one that is based on a sequence of generalized likelihood ratio (GLR) tests (see Section \ref{sec:GLRT}). The proposed approaches are designed specifically for applications with high-dimensional data but small number of samples. They build on a more general technique, which we introduced in \cite{Song2016canonical}, that determines the dimension of the signal subspace correlated between two different data sets. The present letter specializes \cite{Song2016canonical} to the case where the two data sets are $\mathbf{x}$ and its complex conjugate $\mathbf{x}^*$. This, however, is not straightforward and requires special care when counting the number of free parameters in the ITC and deriving the approximating distributions in the hypothesis tests. 

\section{Problem Formulation}

Consider a linear signal-plus-noise model for the generation of the observed data vector $\mathbf{x} \in \mathbb{C}^{m}$ 
\begin{equation}
\mathbf{x} = \mathbf{A} \mathbf{s} + \mathbf{n},
\end{equation}
where $\mathbf{s} \in \mathbb{C}^{d+f}$ is a zero-mean complex Gaussian source vector, $\mathbf{A} \in \mathbb{C}^{{m}\times (d+f)}$  is an unknown but fixed mixing matrix with full column rank, and $\mathbf{n}\in \mathbb{C}^{m}$ is a zero-mean complex Gaussian noise vector independent from the source vector. The following additional assumptions are made: 
\begin{itemize}
\item The source vector contains $d$ improper and $f$ proper signal components. This means that
\begin{align}
&\text{rank} (E[\mathbf{ss}^H]) = d+f ,\notag \\ 
&\text{rank} (E[\mathbf{s(s^*)}^H]) = \text{rank} (E[\mathbf{ss}^T]) = d.
\end{align} 
We also allow $f=0$, i.e., all the signal components may be improper. All signal components are independent, and the dimensions $d$ and $f$ are unknown with $d + f < m$.
\item The noise vector $\mathbf{n}$ is proper and possibly colored with an arbitrary covariance matrix $\mathbf{R}_{nn}$. This is a more general noise model than the one used in \cite{li2011noncircular} where the noise vector is assumed to be white. \end{itemize}
Under the above assumptions, the covariance and the complementary covariance matrices of $\mathbf{x}$ are
\begin{align}
\label{eq:rank_Rxx_tilde}
\mathbf{R}_{xx} &= E[\mathbf{xx}^H] = \mathbf{A} E[\mathbf{ss}^H] \mathbf{A}^H + \mathbf{R}_{nn} , \notag \\
\widetilde{\mathbf{R}}_{xx} &= E[\mathbf{xx}^T]  = \mathbf{A} E[\mathbf{ss}^T] \mathbf{A}^T .
\end{align}
Let us define the complex augmented vector $\underline{\mathbf{x}} = [ \mathbf{x}^T, \mathbf{x}^H]^T$ obtained by stacking $\mathbf{x}$ on top of its complex conjugate $\mathbf{x}^*$. The covariance matrix of $\underline{\mathbf{x}}$ is the augmented covariance matrix \cite{schreier2010statistical}
\begin{equation}
\underline {\mathbf {R}}_{xx} =  E [ \underline {\mathbf {x}} \, \underline {\mathbf {x}}^H] = \left[ 
\begin{matrix}
{\mathbf {R}}_{xx} & \widetilde{\mathbf {R}}_{xx} \\ 
\widetilde{\mathbf {R}}_{xx}^{\ast} & {\mathbf {R}}_{xx}^{\ast} 
\end{matrix}
\right] ,
\end{equation}
which is a convenient way of keeping track of both $\mathbf{R}_{xx}$ and $\widetilde{\mathbf{R}}_{xx}$. In this letter, we are interested in estimating the dimension of the improper signal subspace $d$, which is equal to the rank of $\widetilde{\mathbf{R}}_{xx}$. 

In practice, the true covariance matrices are unknown and have to be estimated from samples. We consider $M$ independent and identically distributed (i.i.d.) samples of $\underline {\mathbf {x}}$, arranged as the $M$ columns of the data matrix $\underline {\mathbf {X}}  = \begin{bmatrix} \underline {\mathbf {x}}(1), \underline {\mathbf {x}}(2), \ldots, \underline {\mathbf {x}}(M) \end{bmatrix} $, where $\underline {\mathbf {x}}(i)$ denotes the $i^{\text{th}}$ sample of $\underline {\mathbf {x}}$. When $\widetilde{\mathbf{R}}_{xx}$ is estimated from $\underline {\mathbf {X}}$, its rank, in general, will not be equal to $d$. In Sections \ref{sec:MDL-ITC} and \ref{sec:GLRT} we will introduce two ways of estimating $d$, which are both based on the \textit{circularity coefficients} of $\mathbf{x}$ \cite{schreier2010statistical}. These are the canonical correlations between $\mathbf{x}$ and $\mathbf{x}^\ast$, which can be computed as the singular values of the coherence matrix $\mathbf{C} =  \mathbf {R}_{xx}^{-1/2} \widetilde{\mathbf {R}}_{xx} \mathbf {R}_{xx}^{-T/2}$. The circularity coefficients are normalized to take values between 0 and 1, and they measure the degree of impropriety of each signal component. A maximally improper component leads to a circularity coefficient of 1, and a proper component to a zero circularity coefficient. When working with samples, the following complication arises. Unless the number of samples is significantly greater than the dimension of the data, the sample circularity coefficients are significantly greater than the (true) population circularity coefficients. As we would like to be able to handle the sample-poor scenario, this requires the use of a dimension-reducing preprocessing step. 

\textit{Note}: Another term common in complex-valued signal processing is \textit{circularity}, which is a stronger version of propriety. For the Gaussian distribution, propriety implies circularity and noncircularity implies impropriety. As we have assumed $\mathbf{x}$ to be zero-mean Gaussian, the improper signal subspace is the noncircular signal subspace. However, in general noncircularity does not imply impropriety \cite{schreier2010statistical}. 

\section{Approach based on Information Theoretic Criterion}
\label{sec:MDL-ITC}
For a given set of observations and a family of models, the ITCs introduced by Akaike \cite{akaike1998information}, Schwarz \cite{schwarz1978estimating} and Rissanen \cite{rissanen1978modeling} select the model that best fits the observation data, while also making sure that the model does not overfit the data. The goodness-of-fit is measured by the likelihood function for $M$ samples of $\underline{\mathbf{x}}$ ,which is parameterized by $\underline {\mathbf {R}}_{xx}$: 
\begin{equation}
f(\underline {\mathbf {X}} | \underline {\mathbf {R}}_{xx} ) = \prod_{i=1}^{M} \frac{1} { \pi^m \sqrt{\det\underline {\mathbf {R}}_{xx}} } \exp{\bigg[-\frac{ \underline{\mathbf{x}}^{H}(i) \underline {\mathbf {R}}_{xx}^{-1} \underline{\mathbf{x}}(i)}{2}\bigg]} .
\end{equation} 
The ITC score is 
\begin{equation}
\label{eq:ITC}
\text{ITC}(d) = -\ln f(\underline {\mathbf {X}} | \hat{\underline {\mathbf {R}}}_{xx}) + \alpha(M)C ,
\end{equation}
where $\hat{\underline {\mathbf {R}}}_{xx}$ is the maximum likelihood estimate of $\underline {\mathbf {R}}_{xx}$ (which is simply the sample augmented covariance matrix), and the second term in \eqref{eq:ITC} is a penalty function that penalizes complex models. In our case, the model order is the number of improper signals, $d$. Both terms in the sum of \eqref{eq:ITC}, i.e. the quality of the fit and the complexity penalty, depend on $d$. In the penalty term, $C$ is the number of free parameters in the parameter space of the model, i.e., in $\underline {\mathbf {R}}_{xx}$. The term $\alpha(M)$ depends on the chosen ITC. We use the MDL criterion as it leads to a consistent estimator of $d$ \cite{wax1985detection}, for which $\alpha(M) = \frac{\ln(M)}{2}$. The MDL-ITC chooses the $d$ that  minimizes (\ref{eq:ITC}), that is 
\begin{equation}
\label{eq:d_ITC}
\hat{d} = \argmin_{d=0,...,m-1}\text{ITC}(d).
\end{equation}
The ITC expression in  (\ref{eq:ITC}) can be simplified as follows.

\textit{Model Fit Score}: The maximization of the log-likelihood is performed under the constraint that $\text{rank}(\widetilde{\mathbf {R}}_{xx}) = d$. The maximum log-likelihood is \cite{schreier2010statistical}
\begin{equation}
\label{eq:likelihood_ITC}
-\ln f(\underline {\mathbf {X}} | \hat{\underline{\mathbf {R}}}_{xx} ) \propto \frac{M}{2} \ln \prod_{i=1}^{d} (1 - \hat{k}_i^2),
\end{equation}
where $\hat{k}_i$ are the sample circularity coefficients of $\mathbf{x}$. 

\textit{Number of Free Parameters}: Since only the complementary covariance matrix of $\mathbf{x}$, $\widetilde{\mathbf {R}}_{xx}$, depends on $d$, only $\widetilde{\mathbf {R}}_{xx}$ instead of the entire $\underline{\mathbf {R}}_{xx}$ is considered when calculating the number of free parameters. To do this, we perform the Takagi factorization for complex symmetric matrices \cite{horn2012matrix} given as 
\begin{equation}
\widetilde{\mathbf {R}}_{xx} = \mathbf{F K F}^T.
\end{equation}
Here, $\mathbf{F}$ is a complex unitary matrix, which contains the singular vectors, and $\mathbf{K} = \text{diag}(k_1, k_2, \ldots, k_d, 0, \ldots, 0)$ contains the $d$ non-zero circularity coefficients. Since $\text{rank}(\widetilde{\mathbf {R}}_{xx}) = d$, there are $2md$ and $d$ free parameters in $\mathbf{F}$ and $\mathbf{K}$, respectively. However, not all of these parameters are freely adjustable. There are $d$ and $d(d-1)$ constraints on the elements of the singular vectors in $\mathbf{F}$ due to normality and orthogonality, respectively. Therefore, 
\begin{align}
\label{eq:C}
C &= 2md + d - (d + d(d-1)) ,\notag \\
   &= 2md - d^2 + d 	.
\end{align}
The simplified MDL-ITC expression is thus given as
\begin{equation}
\label{eq:ITC_samplerich}
\text{ITC}(d) = \frac{M}{2} \ln {\prod_{i=1}^{d} (1 - \hat{k}_i^2)} + \frac{\ln{M}}{2}(2md - d^2 + d).
\end{equation}

\subsection{Sample Poor Scenario} 
\label{sec:ITC_samplepoor}
Unless the number of samples $M$ is significantly larger than the dimension of the observation $m$, the number of improper components $d$ cannot be correctly estimated using (\ref{eq:d_ITC}) because the sample circularity coefficients $\hat{k}_i$ are significantly larger than the population circularity coefficients. Moreover, when $M < 2m$, at least $2m-M$ sample circularity coefficients are 1 independently of the underlying model generating them \cite{pezeshki2004empirical}. This calls for rank reduction before or alongside the estimation of $d$. 

One of the most common rank-reduction methods is PCA. The rank-$r$ PCA description of $\mathbf{x}$ is 
\begin{equation}
\label{eq:PCA}
\mathbf{x} = \mathbf{U}^H_r \mathbf{x} ,
\end{equation}
where $\mathbf{U}_r$ denotes the matrix containing as its columns the first $r$ principal eigenvectors of $\mathbf{R}_{xx}$. Of course, PCA only retains the signal components that have maximum variance within the data. These do not necessarily correspond to the most improper signals, which have maximum covariance between $\mathbf{x}$ and $\mathbf{x}^*$. Nevertheless, following our approach in \cite{Song2016canonical} we can choose $r$ large enough to include all the improper signals, while eliminating much of the noise and those proper components whose variance is smaller than that of the weakest improper component. This can be done based on the following reduced-rank version of the ITC expression in (\ref{eq:ITC_samplerich}): 
\begin{equation}
\label{eq:ITC_samplepoor}
\text{ITC}(d,r) = \frac{M}{2} \ln {\prod_{i=1}^{d} (1 - \hat{k}_i^2(r))} + \frac{\ln{M}}{2}(2rd - d^2 + d).
\end{equation}
The circularity coefficients $\hat{k}_i(r)$ are computed from the rank-$r$ PCA description \eqref{eq:PCA} of the data and thus depend on the rank $r$. They can change significantly depending on how $r$ is chosen. The optimal rank is the one that includes all the improper signal components, but not more than that. \enquote{Detector 3} in \cite{Song2016canonical} allows us to {\em jointly} choose the optimum rank $r$ and estimate the number $d$ of improper components. The decision rule for $d$ is \footnote{While the decision rule \eqref{eq:d_ITC_rr} corresponds to \enquote{Detector 3} in \cite{Song2016canonical}, the expression for $\text{ITC}(d,r)$ in this letter differs from \cite{Song2016canonical} because the number of free parameters are different when analyzing correlation between $\mathbf{x}$ and $\mathbf{x}^*$ rather than two different data sets.}
\begin{equation}
\label{eq:d_ITC_rr}
\hat{d} = \max_{r=1,...,r_{\text{max}}} \argmin_{s=0,...,r-1}\text{ITC}(d,r),
\end{equation}
and the $r$ that leads to $\hat{d}$ is the chosen PCA rank. This decision rule can be motivated as follows. The min-step, which corresponds to the traditional MDL-ITC, generally will not overestimate $d$ because MDL is consistent. However, if $r$ is not chosen large enough, the reduced-rank description will not capture all the improper signals and $d$ could be underestimated. Because the min-step will not overfit, we can simply take the maximum over all $r$ from 1 to $r_{max}$. Here, $r_{\text{max}}$ is the maximum allowable rank and is chosen to be sufficiently smaller than $M$ (typically $M/3$) \cite{Song2016canonical}. This is a much more relaxed condition than requiring $m$ to be sufficiently smaller than $M$. 

\section{Approach based on Hypothesis Testing}
\label{sec:GLRT}

The problem of order selection can also be solved by performing a series of binary hypothesis tests \cite{bartlett1954note,lawley1956tests}. Starting with improper signal counter $s = 0$, each binary test is: 
\begin{align}
H_0 &: \text{{$d=s$}} \notag \\
H_1 &: \text{{$d>s$}}
\end{align}
If $H_0$ is rejected, $s$ is incremented and another test of $H_0$ vs. $H_{1}$ is run. This is repeated until $H_0$ is not rejected or $s$ reaches its maximum possible value. Each binary test is a likelihood ratio test. Since the unknown parameters are replaced by their maximum likelihood estimates, this leads to a generalized likelihood ratio test (GLRT). The GLR for the hypothesis test is
\begin{equation}
\label{eq:lr}
\eta = { { f(\underline {\mathbf {X}}|\hat{\underline{\mathbf {R}}}_{xx},d=s)} \over {  f(\underline {\mathbf {X}}|\hat{\underline{\mathbf {R}}}_{xx}},d>s)} ,
\end{equation}
where $f(\underline {\mathbf {X}}|\hat{\underline{\mathbf {R}}}_{xx},d=s)$ and $f(\underline {\mathbf {X}}|\hat{\underline{\mathbf {R}}}_{xx},d>s)$  are the likelihood functions under the null and the alternative hypothesis, respectively. From (\ref{eq:ITC}) and (\ref{eq:lr}), it can be seen that the GLRT and ITC are related to each other. This has also been shown in \cite{stoica2004information}.  

Since the parameter space for $d=m$ is sufficient to parametrize all the possibilities when $d>s$, we have
\begin{equation}
\label{eq:HTlikelihood2}
 f(\underline {\mathbf {X}}|\hat{\underline{\mathbf {R}}}_{xx},d>s) \propto { \bigg(  \prod_{i=1}^{m} (1 - {\hat{k}_i^2})  \bigg) }^{-\frac{M}{2}} ,
\end{equation}
and thus 
\begin{equation}
\label{GLR}
\eta = \bigg\{ \prod_{i=s+1}^{m} (1 - {\hat{k}_i^2}) \bigg\}^{\frac{M}{2}} .
\end{equation}

According to Wilks' theorem, under $H_0$ the statistic $W(s) = -2 \ln \eta$ is asymptotically $\chi^2$-distributed with degrees of freedom (d.f.) equal to the difference between the numbers of free parameters under $H_{1}$ and $H_{0}$ \cite{wilks1938large}. Under $H_0$, the d.f. are given by (\ref{eq:C}). Under $H_1$, the d.f. are obtained from (\ref{eq:C}) by setting $m = d$. Hence, for $M \rightarrow \infty$, $W(s)$ is $\chi^2$ with $(m - d)(m - d + 1)$ d.f. 

\subsection{Sample Poor Scenario} 
As discussed in Section \ref{sec:ITC_samplepoor}, sample poor scenarios require rank reduction to correctly estimate the number of improper signals. A reduced-rank version of the test statistic $W(s)$ is the Box statistic \cite{box1949general}
 \begin{equation}
B(s,r) = -(M - r)\ln{ \prod_{i=s+1}^{r} \bigg( 1 - {\hat{k}_i^2(r)} \bigg) }, 
\end{equation}
which is approximately $\chi^2$-distributed with $(r-1)(r-d+1)$ d.f. The correction term $(M - r)$ introduced in \cite{walden2009testing} provides a better approximation by the $\chi^2$-distribution than the Wilks statistic for much smaller number of samples. It can be shown numerically that $B(s,r)$ approximately follows a $\chi^2$-distribution as long as $r$ is large enough to capture all the improper components and is also sufficiently small compared to $M$ (as in Section \ref{sec:MDL-ITC}, $r < M/3$ seems to work well). A decision rule can thus be formulated as
\begin{equation}
\label{eq:d_HT_rr}
\hat{d} = \max_{r =1,...,r_{\text{max}}} \min_{s=0,...,r-1} \{ s: B(s,r)  < T(s,r) \} ,
\end{equation}
where $T(s,r)$ is the threshold chosen to maintain a specified probability of false alarm $P_{\text{fa}}$, which can be obtained from the $\chi^2$-approximation. This is \enquote{Detector 1} from \cite{Song2016canonical} specialized to the case of detecting the number of correlated components between $\mathbf{x}$ and $\mathbf{x}^*$. The motivation behind it is similar to the that of \eqref{eq:d_ITC_rr}. While \cite{Song2016canonical} uses a Bartlett-Lawley approximation of the test statistic $W(s)$, the fact that here we are analyzing correlations between $\mathbf{x}$ and $\mathbf{x}^*$ means that the Box statistic with different d.f. needs to be used instead \cite{box1949general}.

\section{Numerical Results}

In this section, we evaluate the performance of the proposed detectors based on ITC and GLRT for the application of sensor array processing. We consider the case when $f = 0$, i.e. the entire signal subspace is improper. This is the scenario used in \cite{charge2001non,haardt2004enhancements,roemer2009multidimensional,delmas2004asymptotically}, which show that utilizing the complementary covariance matrix for direction-of-arrival estimation can lead to significant performance improvement when improper signals such as BPSK-modulated sources impinge on the sensor array. DOA estimation techniques typically assume that the dimension of the signal subspace is known. In practice, this is not the case. If it is known a priori that all sources are improper, then our technique can be employed to find the number of sources. 

The simulation setup is as follows. We use a uniform linear array with $m=60$ sensors with half-wavelength inter-sensor spacing. There are $4$ far-field, narrowband Gaussian sources that impinge on the array at angles $\Theta = [ 10^{\circ},15^{\circ},20^{\circ},25^{\circ}]$. The $q^\text{th}$ column of $\mathbf{A}$ matrix is $[1, \exp{(j\frac{\pi}{2} \cos(\theta_q))}, \ldots,$ $\exp{(j\frac{\pi}{2} (m-1)\cos(\theta_q))}]^T$ for $q=1,\ldots,4$. Each source has variance 5 and the circularity coefficients for the sources are 1, 0.9, 0.8, and 0.6. Two scenarios are presented: a) the additive noise is white and Gaussian distributed with unit variance; b) the noise is filtered through an autoregressive (AR) filter of order 4 and filter coefficients [1/2 $\sqrt7$/4 1/2 1/4]. The variance of noise components before filtering is $1/4$. 

\begin{figure}[!t]
    \centering
    \begin{subfigure}[b]{0.480\textwidth}
        \centering
        \includegraphics[width=.90\textwidth, height=5.5cm]{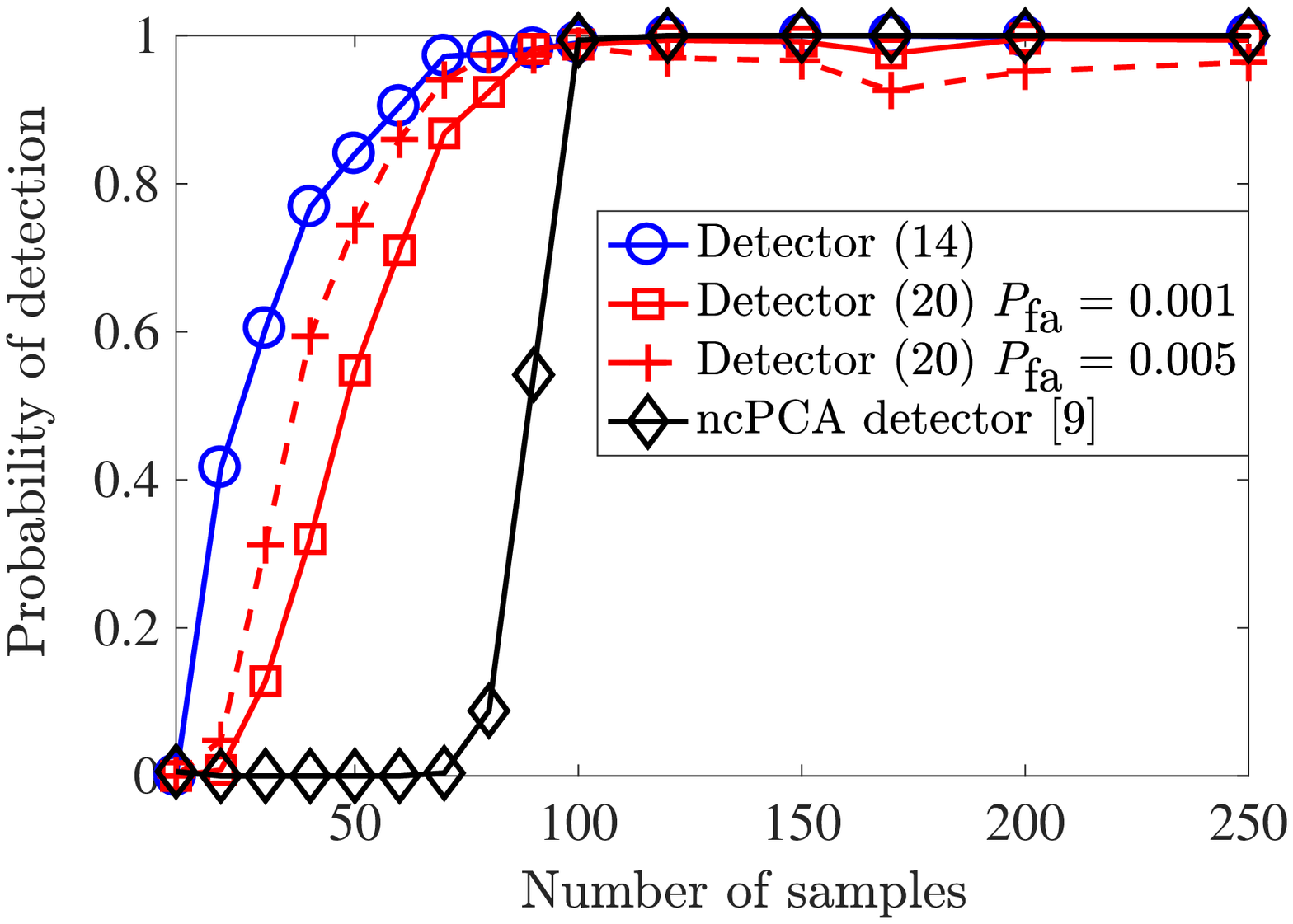}
        \caption{}
        \label{fig:a}
    \end{subfigure}
    \begin{subfigure}[b]{0.480\textwidth}
        \centering
        \includegraphics[width=.90\textwidth, height=5.5cm]{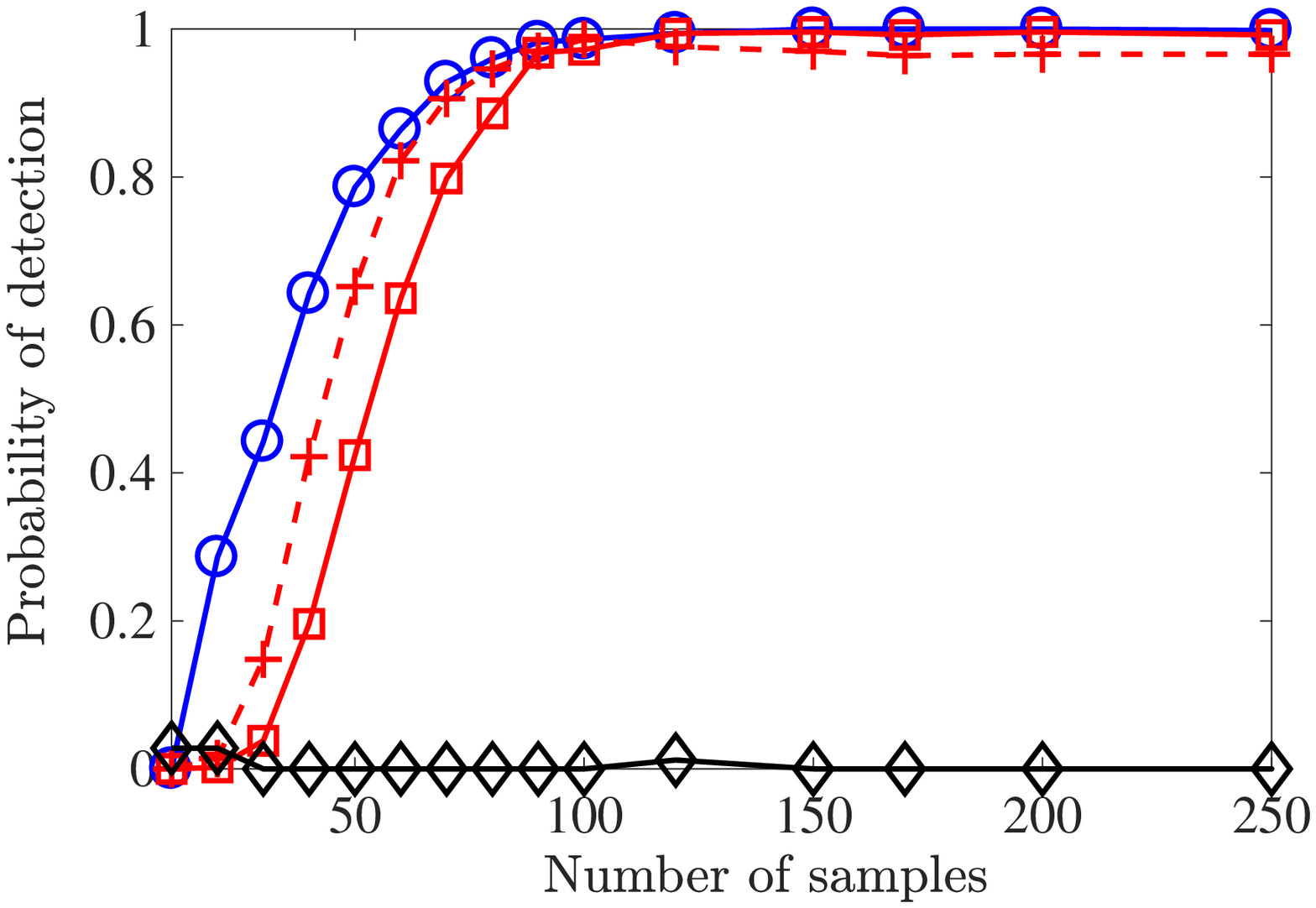}
        \caption{}
        \label{fig:b}
    \end{subfigure}
    \caption{Probability of correctly detecting $d=4$ improper signal components for the proposed detectors and the ncPCA detector in \cite{li2011noncircular} when a) the additive noise is white Gaussian b) the additive noise is colored AR(4). }
    \label{fig:pd_samples}  
    \end{figure}

We compare the performance of our proposed detectors in (\ref{eq:d_ITC_rr}) and (\ref{eq:d_HT_rr}) with the ncPCA detector in \cite{li2011noncircular}. Figure \ref{fig:pd_samples} shows the probability of detection as a function of the number of samples for both scenarios. For each data point, we ran 500 independent Monte Carlo trials. The results for the detector based on a sequence of hypothesis tests are shown for two different values of probability of false alarm, $P_{\text{fa}}$. For the white noise case, all the detectors perform well for a sufficiently large number of samples, but our detectors reach their best performance for smaller number of samples than the ncPCA detector. It is not surprising that the performance of the detector based on hypothesis testing depends on the $P_{\text{fa}}$ value. The detector with $P_{\text{fa}} = 0.005$ performs better than the one with $P_{\text{fa}} = 0.001$ when the number of samples is low. However, when there are enough samples, the detector with smaller $P_{\text{fa}}$ performs better. The variation in performance is due to the fact that a detector with larger $P_{\text{fa}}$ generally tends to overfit, while a detector with smaller $P_{\text{fa}}$ tends to underfit. This requires the right trade-off, which is done automatically in the ITC-based detector.    

In the case of colored noise shown in Figure \ref{fig:b}, the ncPCA detector fails while our detectors continue to work well. This is because the ncPCA detector detects both the proper and improper signal subspaces, and hence must assume white noise in order to distinguish between signal and noise. Since we only identify improper signal components, we only need to assume proper noise, but it does not have to be white.

\section{Conclusion}

We have presented two techniques, based on ITC and hypothesis testing, for detecting the dimension of the improper signal subspace in high-dimensional complex data with additive noise. There is no assumption made on the structure of the covariance matrix of the noise, and we have shown using simulations that the proposed detectors work well even in the presence of colored noise. We have introduced reduced-rank detectors, which work reliably even for small number of samples. 

\bibliographystyle{ieeetran}
\bibliography{references} 

\end{document}